\documentclass[copyright,creativecommons]{eptcs}
\providecommand{\event}{PAR 2010}
\usepackage{latexsym,amssymb,amsfonts,proof,hyperref,breakurl}

\newtheorem{definition}{Definition}[section]
\newtheorem{theorem}[definition]{Theorem}
\newtheorem{lemma}[definition]{Lemma}

\title{General Recursion and Formal Topology}

\def\titlerunning{General Recursion and Formal Topology}
\def\authorrunning{C. Sacerdoti Coen \& S. Valentini}

\author{
Claudio Sacerdoti Coen
\institute{Dipartimento di Scienze dell'Informazione\\
Universisit\`a di Bologna}
\email{sacerdot@cs.unbo.it}
\and
Silvio Valentini
\institute{Dipartimento di Matematica Pura e Applicata\\
Universit\`a di Padova}
\email{silvio@math.unipd.it}
}

\begin{document}
\maketitle

\begin{abstract}
It is well known that general recursion cannot be expressed within Martin-L\"of's type 
theory and various approaches have been proposed to overcome this problem 
still maintaining the termination of the computation of the typable terms.
In this work we propose a new approach to this problem based on the use of
inductively generated formal topologies.
\end{abstract}

%\tableofcontents

\section{Introduction}
\label{introduction}

Martin-L\"of's type theory is at the same time a functional programming language
and a rich specification language which allows the definition of a full intuitionistic logical
calculus since it adheres to the {\em proposition as types} paradigm (see \cite{ML84}).

However, in order to ensure the termination of the computation of every well typed program, 
it does not allow general recursion and hence it does not permit to program
in a natural functional style (see \cite{Hug84}).

On the other hand, each set is defined inductively and hence it is provided with 
a recursion rule which allows the definition of programs by pattern matching on the
possible shapes of an element of that set.
This feature, together with the presence of cartesian products and function space, turns 
type theory into a very flexible programming environment where programs can be 
developed together with the proof of their correctness while termination follows
as a corollary of the head normal form theorem (see \cite{BV92}).

In order to solve the problem of the lack of general recursion still maintaining the 
property of the termination of every well typed program many solutions have been
proposed starting from the first suggestions by Peter Aczel of using an accessibility 
predicate in \cite{Acz77}, the work by Bengt Nordstr\"om where the use of a 
general recursion operator on well ordered set is proposed (see \cite{Nor88}), 
till the more recent papers by Ana Bove and Venanzio Capretta
which suggest to use an {\em ad hoc} accessibility predicate (see \cite{BC05}).

In this paper we want to propose a slightly different approach based
on the use of inductively generated formal topologies that we are going to recall
in the next section.
Indeed, we think that this approach, even if it is just a variation of the one of 
Nordstr\"om and it is less flexible than the one of Bove and Capretta, is offering a much 
better possibility for further development since it is not just a solution for 
a specific problem but it is part of a much deeper mathematical theory: we will 
give some suggestions of these possibilities in the concluding section \ref{developments}.  

Finally, we also show that for the special class of inductively generated
formal topologies that we need for general recursion, it is possible to provide
not only an induction principle, but also a recursion principle for the
representation of the ad-hoc accessibility predicate of Bove and Capretta.
Thus, as for a variant of the Bove-Capretta method, the witnesses of the
accessibility predicate have no concrete computational use and code extraction
yields exactly the same general recursive function that one would write in a
standard functional programming language.

\section{Inductively Generated Formal Topology}
\label{formalTopology}

We are going to recall here only the main definitions on inductively generated 
formal topologies and the results that we need in the following (for a more detail 
account on the topic the reader is invited to refer to \cite{TIG,Val06}).

\begin{definition}[Inductively generated formal topology]
An {\em inductively generated formal topology} is a triple $(A,I,C)$ such that
$A$ is a set, $I(a)$ is a set (of indexes) for any $a \in A$ and $C(a,i)$ 
is a subset of $A$ for any $a \in A$ and $i \in I(a)$.
\end{definition}

In the following we will say that the couple $I(-),C(-,-)$ is an {\em axiom-set} for
the inductively generated formal topology $(A,I,C)$.
\medskip

The name of {\em inductively generated formal topology} for the triple above
is due to the fact that in any inductively generated formal topology $(A,I,C)$ 
it is possible to define an infinitary relation, namely, the {\em formal cover relation} 
$\vartriangleleft$, by using the following inductive rules\footnote{It is worth noting
that we are defining what does it mean to be covered by $U$ and hence the
subset $U$ is not required to appear in the proof term.}
$$
\begin{array}{llll}
\mbox{(reflexivity)}
&\displaystyle{\frac
   {h : a \in U}
   {\mathsf{refl}(a,h) : a \vartriangleleft U}}
&\mbox{(infinity)}
&\displaystyle{\frac
   {i : I(a) \quad k : (\forall y \in C(a,i))\ y \vartriangleleft U}
   {\mathsf{inf}(a,i,k) : a \vartriangleleft U}}
\end{array}
$$
The most direct mathematical interpretation for the elements of $A$ and the formal cover 
relation is respectively into the open subsets of a topological space and its 
coverage relation.
Indeed, in this case, one immediately obtains that if $a \vartriangleleft U$ holds then the interpretation
of $a$ is covered by the open set determined by the union of the open subsets where the elements
of $U$ are interpreted provided the interpretation satisfies the axioms, namely, for any $a \in A$ 
and $i : I(a)$, the interpretation of $a$ is covered by the interpretation of $C(a,i)$ 
(in fact, not only this interpretation is valid, but it is possible to prove that it is also complete when 
a countable number of axioms is considered , see \cite{Val06}).

The previous introduction rules allow an immediate definition of a recursion operator
on the proof terms of the set $a \vartriangleleft U$.
However, to keep the notation simpler we will write the rule without the 
proof-terms\footnote{A stronger version of the induction principle would also be valid. It is obtained by substituting the assumption of the second 
minor premise by
$i : I(x), k : (\forall y \in C(x,i))\ (y \vartriangleleft U) \to P(y)$.
However, the simplified one that we propose here will be 
sufficient to grasp the ideas in the rest of the paper.}
(a complete formalization in the Matita proof assistant can be found in \cite{Sac10}, 
based on \cite{Tas10}).
$$
\mbox{\begin{tabular}{cc}(cover-induction)\\~\end{tabular}}
\quad
\infer
   {P(a)}
   {a \vartriangleleft U
   &
    \deduce[\vdots]
       {P(x)}
       {[x \in U]}
    &
     \deduce[\vdots]
       {P(x)}
       {[i:I(x), (\forall y \in C(x,i))\
       %y \vartriangleleft U \wedge
       P(y)]}}
$$

For instance, this rule allows to prove the following theorem (see \cite{TIG}).

\begin{theorem}
\label{thTransitivity}
Let $(A,I,C)$ be an inductively generated formal topology and $\vartriangleleft$ its cover relation.
Then the following conditions are admissible.
$$
\begin{array}{llll}
\mbox{(axiom cond.)}
&\displaystyle{\frac
   {i \in I(a)}
   {a \vartriangleleft C(a,i)}}
&\mbox{(transitivity)}
&\displaystyle{\frac
   {a \vartriangleleft U \quad (\forall u \in U)\ u \vartriangleleft V}
   {a \vartriangleleft V}}
\end{array}
$$
\end{theorem}
{\bf Proof.}
The {\em axiom condition} is straightforward since by {\em reflexivity} we have that, for any $y \in C(a,i)$,
$y \vartriangleleft C(a,i)$ holds and hence the result follows by {\em infinity}.

On the other hand {\em transitivity} requires a proof by induction on the length of the derivation of 
$a \vartriangleleft U$, namely, {\em cover-induction} has to be used.
Now, if $a \vartriangleleft U$ is generated by {\em reflexivity} because $a \in U$ then $a \vartriangleleft V$ 
follows by logic since we are assuming that, for all $u \in U$, $u \vartriangleleft V$.
On the other hand, if $a \vartriangleleft U$ is generated by {\em infinity} because there exists some 
$i \in I(a)$ such that, for all $y \in C(a,i)$, $y \vartriangleleft U$, then by inductive hypothesis, 
$y \vartriangleleft V$ and hence we can conclude $a \vartriangleleft V$ by {\em infinity}.
\medskip

Besides the previous result, we are going to use only another theorem on a particular class of
inductively generated formal topologies, namely, {\em singleton inductively generated formal topology}, 
that is, an inductively generated formal topology such that, for each $a \in A$, there is exactly one 
axiom, namely, the set of indexes $I(a)$ is a singleton.
From now on, we will deal only with singleton inductively generated formal topologies and this 
is the reason why, for sake of a simpler notation, we will just omit any reference to the set of the 
indexes and its elements and we will say that $(A,C)$ is a singleton inductively generated formal 
topology if $A$ is a set, and $C(a)$ is a subset of $A$ for any $a \in A$.
Of course, also the rules to generate the cover relation are simplified in the obvious way, that is,
$$
\begin{array}{llll}
\mbox{(reflexivity)}
&\displaystyle{\frac
   {h : a \in U}
   {\mathsf{refl}(a,h) : a  \vartriangleleft U}}
&\mbox{(infinity)}
&\displaystyle{\frac
   {k : (\forall y \in C(a))\ y \vartriangleleft U}
   {\mathsf{inf}(a,k): a \vartriangleleft U}}
\end{array}
$$

\begin{theorem}
\label{thSingleForm}
Let $(A,C)$ be a singleton inductively generated formal topology.
If $a \vartriangleleft \emptyset$ then $a \not \in C(a)$.
\end{theorem}
{\bf Proof.}
The proof is by induction on the derivation of $a \vartriangleleft \emptyset$.
Now, $a \vartriangleleft \emptyset$ if and only if, for all $y \in C(a)$, $y \vartriangleleft \emptyset$.
Hence, by inductive hypothesis, $a \vartriangleleft \emptyset$ yields that, for all $y \in C(a)$, 
$y \not \in C(y)$.
Assume now that $a \in C(a)$.
Then, we get $a \not \in C(a)$ by logic and hence we can conclude $a \not \in C(a)$ by 
discharging the assumption.

\section{General Recursion and Formal Topologies}

Now, let us show how singleton inductively generated formal topologies can help in representing
terminating general recursion in Martin-L\"of's type theory.
To begin with we need to illustrate the relation between ordered sets and formal topologies.

\subsection{Ordered Sets and Unary Formal Topologies}
\label{ipergioco}

In this section we want to show that inductively generated formal topologies allow to express relevant
properties of ordered sets.
To this aim let us begin with the following inductive definition.

\begin{definition}[$R$-foundation]
Let $A$ be a set and $R$ be an order relation on elements of $A$.
Then an element $a \in A$ is {\em $R$-founded} if and only if $\neg a R a$ and, for all $x \in A$, 
if $a R x$ then $x$ is {\em $R$-founded}.
\end{definition}

Let us consider now any singleton inductively generated formal topology $(A,C)$ on the set $A$.
Then we have that\footnote{One can consider the inductive rules for the definition of a singleton 
inductively generated formal topology as an equation whose unknown is the set of the
elements covered by $U$; for a general solution of the inductive equation defining a cover relation 
in the case of inductively generated formal topologies see \cite{Val07}.}
$$
a \vartriangleleft U \mbox{ iff } (a \in U) \mbox{ or } (\forall x \in C(a))\ x \vartriangleleft U
$$

Thus, if we instantiate $U$ to the empty set we obtain both that
$$
\begin{array}{lcl}
a \vartriangleleft \emptyset &\mbox{iff} &(\forall x \in C(a))\ x \vartriangleleft \emptyset
\end{array}
$$
and, by Theorem \ref{thSingleForm}, that $a \not \in C(a)$.

So, if $(A,R)$ is an ordered set and $C(a) \equiv \{ x \in A \mid a R x \}$ we arrive at 
the following statement (for a deeper analysis of the situation and a complete proof of
the next theorem showing the connection between inductively generated formal topologies
and the tree set in \cite{NPS90} the reader is invited to look at \cite{Val07,Val10}).

\begin{theorem}
\label{thMain}
Let $(A,R)$ be an ordered set and put, for any $x \in A$, $C_R(x) \equiv \{ y \in A \mid x R y \}$.
Now, suppose that $a$ is an element of $A$, then
$a$ is $R$-founded if and only if $a \vartriangleleft_R \emptyset$ in the singleton inductively
generated formal topology $(A,C_R)$.
\end{theorem}

\subsection{Implementing Terminating General Recursion}

The general shape of a functional program $f$ on elements of a domain $A$ that uses 
\emph{simple general recursion} --- that is, recursive nested calls are not
allowed, the recursively defined function is always fully applied and never
passed to higher order functions, see~\cite{BC05} --- can be represented as
$$
\begin{array}{lcl}
f(x_1) &= &g_1(f(d_{1,1}(x_1)), \ldots, f(d_{1,n_1}(x_1)))
\\
\ldots
\\
f(x_k) &= &g_k(f(d_{k,1}(x_k)), \ldots, f(d_{k,n_k}(x_k)))
\end{array}
$$ 
where $x_1,\ldots,x_k$ are all the disjoint possible shapes of the elements of $A$.

So the computation of $f$ on an element of $A$ gives rise to a finitary computation tree which can 
be either finite or not.

Thus, we are naturally led to define an order relation $R_f$ between elements of $A$ such that the 
element $x_i \in A$ is related with all those elements $d_{i,1}(x_i)$, \ldots, $d_{1,n_i}(x_i)$ whose 
evaluation by $f$ is necessary in order to evaluate $f$ on $x_i$.

Moreover, it is clear that the computation of $f$ on $x_i$ is terminating if and only if it is terminating
on all the elements $d_{i,1}(x_i)$, \ldots, $d_{1,n_i}(x_i)$, namely, if and only if the element
$x_i$ is $R_f$-founded.

Then, after Theorem \ref{thMain}, we get that the computation of $f$ on $x_i$ is terminating if and only if
$x_i \vartriangleleft_{R_f} \emptyset$.
Thus, we can define a functional program that emulates the computation of $f$ on the terminating values
by structural recursion on the proof of $x_i \vartriangleleft_{R_f} \emptyset$.

Let us illustrate our statement on a simple example.

\subsubsection{The Fibonacci Function}

The standard general recursive definition of the Fibonacci function is
$$
\begin{array}{lcl}
\mathsf{fib}(0) &= &0
\\
\mathsf{fib}(1) &= &1
\\
\mathsf{fib}(x+2) &= &\mathsf{fib}(x+1) + \mathsf{fib}(x)
\end{array}
$$
It is easy to find a solution for programming this function in standard type theory by exploiting the possibility
to define cartesian products.

However, we are here interested in showing how to use singleton inductively generated formal topologies
to implement the Fibonacci function.

Thus, let us consider the relation $R_{\mathsf{fib}}$ such that $0$ and $1$ are related to no element
while every natural number $x+2$, greater or equal to $2$, is related to $x+1$ and $x$.
According to the abstract analysis above, we induce from this relation the axiom-set 
$C_{R_{\mathsf{fib}}}(0) = \emptyset$, $C_{R_{\mathsf{fib}}}(1) = \emptyset$ and 
$C_{R_{\mathsf{fib}}}(x+2) = \{ x+1, x \}$.
So, we get that $\mathsf{inf}(0,h)$ is a proof element for $0 \vartriangleleft_{R_{\mathsf{fib}}} \emptyset$
if $h$ is the proof element for $(\forall x \in \emptyset)\ x \vartriangleleft_{R_{\mathsf{fib}}} \emptyset$,
which clearly exists after {\em ex falsum quodlibet}; in a similar way $\mathsf{inf}(1,k)$, where $k$ is a 
proof element for $(\forall x \in \emptyset)\ x \vartriangleleft_{R_{\mathsf{fib}}} \emptyset$, is a proof 
element for $1 \vartriangleleft_{R_{\mathsf{fib}}} \emptyset$  and $\mathsf{inf}(x+2,m)$, where $m$ is a 
proof element for $(\forall y \in \{ x+1,x \})\ y \vartriangleleft_{R_{\mathsf{fib}}} \emptyset$, is a proof 
element for $x+2 \vartriangleleft_{R_{\mathsf{fib}}} \emptyset$ .

Thus, we can define the Fibonacci function by structural recursion on the proof element for 
$x \vartriangleleft_{R_{\mathsf{fib}}} \emptyset$ by setting
$$
\begin{array}{lcl}
\mathfrak{fib}(0,\mathsf{inf}(0,h)) &= &0
\\
\mathfrak{fib}(1,\mathsf{inf}(1,k)) &= &1
\\
\mathfrak{fib}(x+2,\mathsf{inf}(x+2,m)) &= &\mathfrak{fib}(x+1,m(x+1)) + \mathfrak{fib}(x,m(x))
\end{array}
$$

Then, if we are able to prove that, for any natural number $x$, 
$x \vartriangleleft_{R_{\mathsf{fib}}} \emptyset$ holds
we will get a proof of the termination of the Fibonacci function on every natural number,
but it is worth noting that we can still define the function even if we do not know that it is
terminating on all natural numbers.

\section{Recursion on the Cover Predicate}
\label{recursion}
So far we have been quite informal on the exact flavour of type theory we
are working in. In particular, if we assume Martin-L\"of's intuitionistic type 
theory, fully embracing the proof-as-types paradigm, we can make no
distinction between the universe of propositions and that of types. Hence
we can identify subsets of $A$, which are functions from $A$ to the universe
$\textbf{Prop}$ of propositions, with families of types indexed over $A$.
Under this identification, the cover-induction principle defined in
Section~\ref{formalTopology} can be applied both to prove that an element
$a$ belongs to a set $V$ or to build an inhabitant of the data type $V(a)$.
The latter usage is the one that justified the definition of the Fibonacci
function given in the previous section.

Other versions of type theory, like the Calculus of (Co)Inductive
Constructions implemented in Coq~\cite{Coq8.2} and Maietti's Minimal Type
Theory~\cite{Mai09}, depart from Martin-L\"of's tradition by clearly
distinguishing propositions from types by separating them into different
universes.
The separation is reflected in the separation between induction and recursion:
given a proof term $p$ for a predicate $P$, it is allowed to prove another
predicate $Q$ by induction over $p$, but not to inhabit a data type by recursion
over $p$. Both induction and recursion are allowed instead when $p$ is an
inhabitant of a data type. The reader can consult~\cite{Mai09} for some
motivations for this restriction. We just recall that, in the restricted
setting, proof terms have no role in the computation of functions and thus
that all propositions are identified with the unit data type during code
extraction (see~\cite{Let08}), yielding more efficient code. Moreover, since
the proof terms are ignored by code extraction, the use of classical logic and,
more generally, of axioms that break cut elimination, does not jeopardise
computability of the extracted function.

For the rest of this section we assume to be in the restricted version of type
theory and we note that the cover-induction principle cannot be applied as it
is to obtain a representation of general recursive functions, unless we
artificially replace the cover predicate $a \vartriangleleft_{R} \emptyset$
with an isomorphic data type, losing all the benefits of the distinction
between proofs and types (and doubling the constant for the computational
complexity of the extracted code, since the inhabitant of
$a \vartriangleleft_{R} \emptyset$ would be computed as well).
Instead, we note that the following recursion
principle for singleton inductively generated formal topologies can be added
without breaking logical consistency:

$$
\mbox{\begin{tabular}{cc}(cover-recursion)\\~\end{tabular}}
\quad
\infer
   {T(a)}
   {a \vartriangleleft \emptyset
   &
     \deduce[\vdots]
       {T(x)}
       {[(\forall y \in C(x))\ T(y)]}}
$$

The cover-recursion principle can actually be defined in the modern versions
of the Calculus of (Co)Inductive Construction whose primitive operators are
well founded structural recursion (which is not restricted by the proof vs
types separation) and case analysis (which is restricted by not allowing to
perform case analysis over a proof term to inhabit a data type,
see~\cite{Coq8.2}). The following proof term has been formalised in~\cite{Sac10}
in the Matita interactive theorem prover~\cite{ASTZ07} and it type-checks
according to the rules presented in~\cite{ARST09}:

$$
\begin{array}{l}
\textbf{let rec}~\textrm{cover-recursion}_{T,H}(a,p:a \lhd \emptyset) : T(a) :=\\
~~~H(a,\lambda y \in C(a). \textrm{cover-recursion}_{T,H}(y,\pi(a,p,y)))\\
\textbf{where}\\
~~~\pi(a,\mathsf{refl}(a,h: a \in \emptyset),y) : y \lhd \emptyset := \textrm{ex-falso}(h)\\
~~~\pi(a,\mathsf{inf}(a,h: (\forall x \in C(a)) x \lhd \emptyset),y) : y \lhd \emptyset := h(y)
\end{array}
$$
where $T: A \to Type$ (i.e. $T$ is a family of types indexed over $A$)
and $H$ is the higher order function $H: (\forall x:A)(\forall y \in C(x))T(y) \to T(x)$.

As a comparison, the canonical proof term automatically generated for the
cover-induction principle is:
$$
\begin{array}{l}
\textbf{let rec}~\textrm{cover-induction}_{P,H}(a,p:a \lhd \emptyset) : P(a) :=\\
~~~\textbf{match}~p~\textbf{with}\\
~~~~~\mathsf{refl}(a,h: a \in \emptyset) \Rightarrow \textrm{ex-falso}(h)\\
~~~~~\mathsf{inf}(a,h: (\forall x \in C(a)) x \lhd \emptyset) \Rightarrow
H(a,\lambda y \in C(a). \textrm{cover-recursion}_{P,H}(y,h(y)))\\
\end{array}
$$
where $T: A \to Prop$ (i.e. $T$ is a predicate),
$H: (\forall x:A)(\forall y \in C(x))P(y) \to P(x)$
(i.e. $H$ is an ordinary induction hypothesis) and the pattern matching
is an ordinary proof by cases.

The idea for the cover-recursion principle, which is not novel (see \cite{BC04}),
and has been already applied in similar form
to the Bove-Capretta method, consists in noting that it is possible to
immediately perform the recursive call before doing case analysis over the
proof term $p$ (which is done by the $\pi$ function). Indeed, all the
computational arguments of the function call (which is just $y$ in our case) can
be discovered without inspecting $p$, which is required only to inhabit the last
argument which is a proof term (and thus can be computed by induction over $p$).
Note, however, that this technique is rarely exploitable. For instance, it
cannot be applied in any way for the general case of non singleton inductively
generated formal topologies since, in that case, we would have to guess the
index $i \in I(a)$ to perform the recursive call on without inspecting the
proof term $p$ which hides $i$. This constitute further evidence that singleton
generated formal topologies are the natural subclass of formal topologies
that is inherently linked to general recursion.

The first branch of the $\pi()$ function obtains a proof of $y \lhd \emptyset$
by ex-falsum, from the assumption that $U = \emptyset$ is inhabited. Hence the
reader can imagine that restricting the cover relation to the case
$U = \emptyset$ is necessary. Actually, the recursion principle can be extended
to deal with any decidable set $U$, i.e. for any set $U$ such that membership
to $U$ is decidable. We explored this solution in~\cite{Sac10}, but it turned
out that for any singleton axiom-set $(A,C)$ and for every $U \subseteq A$
there exists another axiom set $(A,C')$ such that $a \lhd_{(A,C)} U$ iff
$a \lhd_{(A,C')} \emptyset$ (see section \ref{genericCovering}). 
%Intuitively, a general recursive function defined
%by recursion over a proof of $a \lhd U$ behaves like the same function defined
%by recursion over $a \lhd \emptyset$ but for the cases where $a \in U$ that can
%be freely overridden. The corresponding axiom set $(A,C')$
%augments the base cases of the function with the overridden cases in $U$,
%so that the overriding set $U$ can be dropped.

Applying the code extraction procedure of~\cite{Let08} to our proof of the
cover-recursion principle we obtain the following ML-like code which is
clearly the most general implementation of a simple general recursion function
whose associated functional is $H$:

$$
\begin{array}{l}
\textbf{let rec}~\textrm{cover-recursion}_{H}(a) : T(a) :=\\
~~~H(a,\lambda y. \textrm{cover-recursion}_{H}(y))
\end{array}
$$

For our Fibonacci example, after code extraction $H$ is defined in the
expected way

$$
\begin{array}{lcl}
H_\mathsf{fib}(0,\_) &= &0
\\
H_\mathsf{fib}(1,\_) &= &1
\\
H_\mathsf{fib}(x+2,f) &= &f(x+1) + f(x)
\end{array}
$$

\noindent
and the recursive Fibonacci's function is defined simply as

$$
\mathsf{fib} := \textrm{cover-recursion}_{H_\mathsf{fib}}
$$

\section{Further Developments}
\label{developments}
It is clear that our approach is not much different from the one already proposed in \cite{BC05}
or even \cite{Nor88}.
However, substituting a general theory for an {\em ad hoc} one could shed more light or allows 
some generalizations.
%From the topological point of view,
%our technique consists in associating to each general recursive function
%a basic topology, i.e. a generalization of a formal topology where convergence
%is dropped (see~\cite{Sam10} for a thorough development of basic topology).
%We look now at the interpretation of the most simple ``topological'' concepts
%of basic topology to understand in what sense our approach could be more
%informative than the standard one.

In this concluding section we want to suggest some possible developments in this direction.
In order to get them we will exploit the topological meaning of many of the concepts that
we introduced.

\subsection{Generalizing the Cover}

The first kind of generalizations that we can suggest concern the cover relation.
There are many development directions here: first of all one can consider the possibility to cover by a 
generic set instead of an empty one, then we can consider the case the subset $C(a)$ covering by
axiom the element $a$ is non finite, ad finally we can consider non-singleton formal topologies.

\subsubsection{Covering by a Generic Set}
\label{genericCovering}

Till now in developing our proposal we always considered the notion of ``being
covered by the empty set''.
This is due to the fact that in this way we can recover the meaning of a
generic accessibility predicate; however, in formal topology we can as well consider the
notion of being covered by a set $U$.
Thus, in this section we will drop the restriction to the case $a \vartriangleleft \emptyset$ 
and we will consider general covers of the form $a \vartriangleleft U$.

In order to understand what the consequences of this generalization are let
us analyze the meaning of $a$ being covered by $U$.
It means that the computation of $f(a)$ is {\em barred} by $U$ (see \cite{Val07}), namely, that 
every branch in the tree of recursive calls of $f$ rooted in $f(a)$ eventually passes through 
some $f(x)$ for $x \in U$. 
In particular, $a \vartriangleleft \emptyset$ means that the computation tree rooted in $f(a)$ 
is finite, that is, the function $f$ converges on $a$.
But it also means that $f(a)$ is computable under the assumption that $f(x)$ is
computable for every $x \in U$. 
Hence the cover relation captures the notion of {\em relative computability}.

Moreover, when $U$ is decidable and we actually know the value of $f(x)$ for each $x \in U$, 
then we can use this knowledge to compute $f(a)$; indeed, it is sufficient to change $f$ so that 
it first checks if its input is in $U$ and in this case it stops immediately with no need for the 
recursive calls.
This could probably be used to force a diverging function to converge on some inputs 
by stopping the computation on $U$ and returning some value.

This simple consideration can be given the shape of an abstract result.

\begin{lemma}
Let $(A,C)$ be a singleton inductively generated formal topology and $U$ be a
decidable subset of $A$.
Then let us set
$$
C'(a) =
\left \{
\begin{array}{ll}
\emptyset &\mbox{if $a \in U$}
\\
C(a) &\mbox{otherwise}
\end{array}
\right .
$$
and consider the singleton inductively generated formal topology $(A,C')$.
Then $a \vartriangleleft U$ if and only if $a \vartriangleleft' \emptyset$.
\end{lemma}
{\bf Proof.}
In both directions the proof is by induction on the length of the derivation.
So let us suppose that $a \vartriangleleft U$ in order to show that
$a \vartriangleleft' \emptyset$.
Now, if $a \vartriangleleft U$ because $a \in U$ then $C'(a) = \emptyset$
and hence $(\forall y \in C'(a))\ y \vartriangleleft' \emptyset$ holds and so
$a \vartriangleleft' \emptyset$ follows by {\em infinity}.
And, if $a \vartriangleleft U$ because, for any $y \in C(a)$, $y \vartriangleleft U$
then by inductive hypothesis, for any $y \in C(a)$, $y \vartriangleleft' \emptyset$
which yields that, for any $y \in C'(a)$, $y \vartriangleleft' \emptyset$, since 
$C'(a) \subseteq C(a)$, and thus $a \vartriangleleft' \emptyset$ follows by {\em infinity}.

On the other hand, if $a \vartriangleleft' \emptyset$ then, for any $y \in C'(a)$, 
$y \vartriangleleft' \emptyset$, and hence, for any $y \in C'(a)$, $y \vartriangleleft U$
by inductive hypothesis; now let us argue by cases according to membership of $a$ to $U$:
if $a \in U$ then $a \vartriangleleft U$ follows by {\em reflexivity} and otherwise
$C(a) = C'(a)$ and hence $a \vartriangleleft U$ follows by {\em infinity}.
\medskip

From the point of view of an accessibility predicate this just mean that if we have a relation
$R$ then $a$ is covered by $U$ in the singleton inductively generated formal topology
$(A,C_R)$ if and only if $a$ is $R'$-founded for the relation $R'$ such that
$a R' x$ if $a R x$ and $a \not \in U$ and $\{ x \in A \mid a R' x \} = \emptyset$ if $a \in U$.
\medskip

%\textbf{the saturation of $U$ ($\vartriangleleft U$):} the saturation of $U$ is
%defined as $\{a | a \vartriangleleft U\}$. It is the largest set such that
%$x \in \vartriangleleft U$ iff $f(x)$ is computable relatively to $U$.
%In particular, $\vartriangleleft \emptyset$ is the domain of $f$.
%
%~\\
%
%\textbf{formal open ($U = \vartriangleleft U$):} a formal open is a set equal
%to its saturation. It means that $f(x)$ where $x \in U$ either converges or it
%is barred by $U$, i.e. the function either converges or it diverges recursively
%calling itself only on inputs in $U$. The latter condition is a more informative
%definition of divergence, since it describes the behaviour of the diverging
%function. However, divergence is better grasped by the positivity predicate
%(see below). Hence, the notation of formal open is the less informative one in
%our setting.

\subsubsection{Generalizing to Infinite Axioms}

In all the possible examples of an axiom-set obtained from a certain function $f$ 
the set $C_{R_{\mathsf{f}}}(a)$ is always finite for any $a \in A$.
But, singleton inductive generated formal topologies support a more general definition 
which admits any kind of subset of $A$ in the axioms.
So the open problem is: which kind of computable functions can take advantage of 
an infinite subset?

The answer should pass through a formalism which allows a function to have an infinite
amount of arguments, but which is still computable, since we should have a clause like
$$
f(x) = g(f(d_1(x)),f(d_2(x)), \ldots)
$$
For instance, one can suppose that $g$ is able to provide some more output by using 
some more of its infinite amount of arguments, namely, $g$ should be a continuous function. 

From a topological point of view, here it can be useful the notion of compactness
of an element $a$: $a$ is compact if, whenever $a \vartriangleleft U$,
there exists some finite subset $V$ of $U$ such that $a \vartriangleleft V$. 
This means that, if the calling tree of $f(a)$ is barred by an infinite subset $U$ 
(and hence it can potentially require infinitely many recursive calls), then it is also barred
by a finite $V$, that is, the function can be rewritten in such a way that the tree becomes
finitely branching. 

This notion needs more investigation and is linked to the previous ideas of recursive 
functions that perform an infinite number of recursive calls. 
It seems to capture those that are actually computable.

\subsubsection{The Case of the Non-Singleton Formal Topologies}

If we drop the restriction to singleton inductively generated formal topologies,
but we keep the same intuition, then we obtain non-deterministic functions
that, given an input $x$, can perform different sets $C(a,i)$ of recursive
calls for each $i \in I(a)$, possibly yielding different results.

For example, the axiom-set of the non deterministic function
$$
\begin{array}{lcl}
f(0) &= &0 \\
f(1) &= &1 \\
f(n+2) &= & f(n)~|~f(n+1)
\end{array}
$$
that can call either $f(n)$ or $f(n+1)$ when the input is $f(n+2)$, would be\\

\begin{tabular}{l@{ }l@{ }ll@{ }l@{ }ll@{ }l@{ }l}
$I(0)$ & = & $\{0\}$ & $I(1)$ & = & $\{0\}$ & $I(n+2)$ &=& $\{0,1\}$\\
$C(0,0)$ & = & $\emptyset$ & $C(1,0)$ & = & $\emptyset$ &
$C(n+2,0)$& = &$\{n\}$\\ &&&&&& $C(n+2,1)$ &=& $\{n+1\}$
\end{tabular}

This clearly extends the Bove-Capretta approach, but it also requires a precise 
definition of the formalism for non deterministic functions.
Moreover, the cover-recursion principle that we provided for the type theories 
that separate propositions from types only applies to singleton generated 
formal topologies, but, for the case of an enumerable set of indexes $I$, 
we expect to be able to write a similar principle that computes the set of 
all possible non-deterministic outcomes of the reduction process.

\subsection{The Role of the Positivity Predicate}

We can exploit also other features offered by formal topology if we recall that inductively generated 
formal topologies, apart for the inductive definition of the cover relation that we already recalled, allow also 
the definition of a positivity predicate by using the following co-inductive rules:
$$
\begin{array}{llll}
\mbox{($\ltimes$-reflexivity)}
&\displaystyle{\frac
   {a \ltimes F}
   {a \in F}}
&\mbox{($\ltimes$-infinity)}
&\displaystyle{\frac
   {a \ltimes F \quad i \in I(a)}
   {(\exists y \in C(a,i))\ y \ltimes F}}
\end{array}
$$

The intended topological meaning of the positivity predicate $a \ltimes F$ is that the basic open where
$a$ is interpreted meets, namely, has inhabited intersection, with the close set determined by
$F$ whose points are all the $\alpha$ such that if $\alpha$ is contained in the interpretation
of $b$ then $b \in F$ (see \cite{Sam02}).

Like with the cover relation also here we can greatly simplify the co-inductive rules if we
consider the case of a singleton inductively generated formal topology and we instantiate 
$F$ on the whole set $A$.
Indeed, we get that the only relevant rule is a simplified version of {\em $\ltimes$-infinity}
$$
\mbox{($\ltimes$-infinity)}
\quad
\displaystyle{\frac
   {a \ltimes A}
   {(\exists x \in C(a))\ x \ltimes A}}
$$

If we consider now the axioms set $C_{R_{\mathsf{f}}}$, defined after the relation $R_{\mathsf{f}}$
for some function $f$, then we get that an element $a \in A$ is positive with $A$ if and only if
the computation of the function $f$ is not terminating on $a$ since in order to have that $a \ltimes A$
holds we need an infinite $R_{\mathsf{f}}$-chain $a R_{\mathsf{f}} x_1 R_{\mathsf{f}} x_2 \ldots$
such that $x_1 \in C_{R_{\mathsf{f}}}(a)$, $x_2 \in C_{R_{\mathsf{f}}}(x_1)$, \ldots. 
The proof is simply a direct application of the following co-induction principle for
$\ltimes$ to the predicate $P$ stating the existence of the infinite chain.
$$
\mbox{\begin{tabular}{cc}(positivity-coinduction)\\~\end{tabular}}
\quad
\infer
   {a \ltimes F}
   { P(a)
   & \deduce[\vdots]
      {(\exists x \in C(b)) P(x)}
      {[P(b)]}}
$$

More generally, $a \ltimes F$ means that $f(a)$ is
diverging and that all recursive calls are recursively made on elements of $F$
only. 
It is an informative (or positive) definition of divergence since it tells
us how the function diverges (to be compared to the
negative definition ``non converging''). 
Since we know how the function diverges, we can exploit this information, for 
instance to monitor the amount of memory that will be used in the computation. 
The less informative use is the one we presented above, that is, $a \ltimes A$: 
it just says that $f(a)$ diverges (since $A$ is the set of all values).
More generally, the positivity predicate tends to capture liveness properties of 
processes (see~\cite{HH06}).

%~\\
%
%\textbf{the reduction of $F$ ($\ltimes F$):} the reduction of $F$ is
%$\{x | x \ltimes F\}$. It is the smallest subset of $F$ such that for every
%$a$ in $\ltimes F$, $f(a)$ diverges and every recursive call is done on some
%$x$ in $\ltimes F$. Reducing $F$ fleshes out from $F$ the inputs where the
%function diverges respecting some behaviour.
%
%~\\
%
%\textbf{formal close ($F = \ltimes F$):} a formal close is a set equal
%to its reduction. It means that $f(x)$ where $x \in F$ diverges and every
%recursive call is done on some $y$ in $F$. It is a closed set that captures
%some behaviour of a divergent function. The difference with the notion of
%formal open is that $f(x)$ cannot converge for each $x \in F$.

\subsection{Further Research Directions}

Other topological concepts are likely to be informative as well. 
In particular, it would
be interesting to consider real formal topologies, i.e. basic topologies with
convergence. They are obtained by adding a partial order $\leq$ over basic opens
(usually meaning that $a \leq b$ if and only if $a$ is more informative than $b$) and
asking the cover relation to respect this order (see~\cite{TIG}). In our case
the order must be a partial order over the possible inputs of the function $f$.
In particular, we would obtain properties such as: if $a \leq b$ and
$b \vartriangleleft U$ then $a \vartriangleleft U$, meaning that if $b$ is
computable relatively to $U$, then $a$ also is (but not requiring $f(a)$ to
perform a single recursive call on $f(b)$). Hence it could have applications to
the study of relative computability.

It would also be interesting to try to extend the proposed approach to non simple
general recursion. In particular, as for the Bove-Capretta method, in order
to capture nested recursion the most natural way would be to use
induction-recursion to simultaneously define the axiom set together with a
general recursive function given by recursion over a proof that
$a \lhd \emptyset$ where the cover relation is determined by the axiom set
under definition. At the moment, as far as we know,
inductively-recursively generated formal topologies have never been considered in
the literature and it is unknown if they capture more examples of formal
topologies and if interesting examples are among the captured ones.

\section{Conclusion}
We have shown that to each general recursive function $f$ we can associate a
basic topology that describes its domain. In particular, standard topological
notions (like cover, positivity, that is the dual of cover, compactness, etc.)
become informative characterizations of the domain of $f$. Moreover, in
type theory $f(x)$ can be actually defined by recursion over the covering
predicate $x \vartriangleleft \emptyset$ in such a way that the code obtained
by proof extraction is the naive general recursive description of $f$.

So far, our technique does not enlarge the class of general recursive functions
that can be already described in type theory using Nordstrom's well-founded
recursion or Bove-Capretta's method. In particular, some variants of Bove-Capretta's 
method even capture more functions. However, we believe that our work could
help sheding more light on the topological content of the above methods and
suggest more informative proof and representation techniques. For instance, the
positivity predicate can be used to characterize the behaviour of divergent
computation, i.e. its liveness properties (when the computation is supposed to
diverge). It also naturally points to the investigation of different models
of computation, like non determinism or non finitely branching recursion.

\bibliographystyle{eptcs}

\end{document}